\documentclass[seceq]{ptptex}

\usepackage{graphicx}
\usepackage{wrapft}

\usepackage{epsfig,latexsym,amssymb}
\usepackage{amsmath}

\def\eq#1{{Eq.~(\ref{#1})}}
\def\fig#1{{Fig.~\ref{#1}}}
\def\peq#1{{(\ref{#1})}}

\title{Shock Wave Collisions and Thermalization in AdS$_5$}

\author{
Yuri V. \textsc{Kovchegov}%
}

\inst{
Department of Physics, The Ohio State University, Columbus,
OH 43210, USA
}

\abst{
  We study heavy ion collisions at strong 't Hooft coupling using
  AdS/CFT correspondence. According to the AdS/CFT dictionary heavy
  ion collisions correspond to gravitational shock wave collisions in
  AdS$_5$. We construct the metric in the forward light cone after the
  collision perturbatively through expansion of Einstein equations in
  graviton exchanges.  We obtain an analytic expression for the metric
  including all-order graviton exchanges with one shock wave, while
  keeping the exchanges with another shock wave at the lowest order.
  We read off the corresponding energy-momentum tensor of the produced
  medium. Unfortunately this energy-momentum tensor does not
  correspond to ideal hydrodynamics, indicating that higher order
  graviton exchanges are needed to construct the full solution of the
  problem.  We also show that shock waves must completely stop almost
  immediately after the collision in AdS$_5$, which, on the field
  theory side, corresponds to complete nuclear stopping due to strong
  coupling effects, likely leading to Landau hydrodynamics.  Finally,
  we perform trapped surface analysis of the shock wave collisions
  demonstrating that a bulk black hole, corresponding to ideal
  hydrodynamics on the boundary, has to be created in such collisions,
  thus constructing a proof of thermalization in heavy ion collisions
  at strong coupling. }

\PTPindex{228,121}  

\begin{document}

\maketitle

\section{General Setup: Expansion in Graviton Exchanges}
\label{general}

AdS/CFT correspondence conjectures that the dynamics of ${\cal N} =4$
$SU(N_c)$ SYM theory in four space-time dimensions is dual to the type
IIB superstring theory on AdS$_5 \times$S$_5$ \cite{Maldacena:1997re}.
In the limit of large number of colors $N_c$ and large 't Hooft
coupling $\lambda = g^2 N_c$ (with $g$ the gauge coupling constant)
such that $N_c \gg \lambda \gg 1$, AdS/CFT correspondence reduced to
the gauge-gravity duality: ${\cal N} =4$ SU($N_c$) SYM theory at $N_c
\gg \lambda \gg 1$ is dual to (weakly coupled) classical supergravity
in AdS$_5$. Hence the gauge theory dynamics at strong coupling, which
includes all-orders quantum effects, is equivalent to the classical
dynamics of supergravity. Instead of summing infinite classes of
Feynman diagrams in the gauge theory or using other non-perturbative
methods, one can simply study classical supergravity in 5 dimensions.
For a review of AdS/CFT correspondence see \cite{Aharony:1999ti}.

\begin{figure}
  \begin{center}
    \includegraphics[width=5.cm]{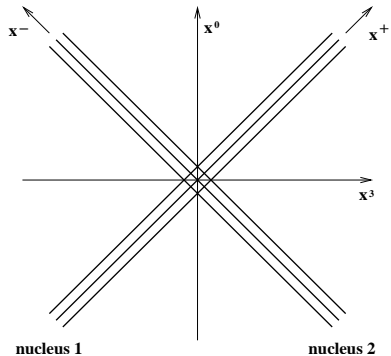}
  \end{center}
  \caption{The space-time picture of the ultrarelativistic heavy ion 
    collision in the center-of-mass frame. The collision axis is
    labeled $x^3$, the time is $x^0$.}
  \label{spacetime}
\end{figure}

Our goal is to describe the isotropization (and thermalization) of the
medium created in heavy ion collisions assuming that the medium is
strongly coupled and using AdS/CFT correspondence to study its
dynamics. We want to construct a metric in AdS$_5$ which is dual to an
ultrarelativistic heavy ion collision as pictured in \fig{spacetime}.
Throughout the discussion we will use Bjorken approximation of the
nuclei having an infinite transverse extent and being homogeneous (on
the average) in the transverse direction, such that nothing in our
problem would depend on the transverse coordinates $x^1$, $x^2$
\cite{Bjorken:1982qr}.

We start with a metric for a single shock wave moving along a light
cone \cite{Janik:2005zt}:
\begin{equation}\label{1nuc}
  ds^2 \, = \, \frac{L^2}{z^2} \, \left\{ -2 \, dx^+ \, dx^- + \frac{2
      \, \pi^2}{N_c^2} \, \langle T_{--} (x^-) \rangle \, z^4 \, d
    x^{- \, 2} + d x_\perp^2 + d z^2 \right\}.
\end{equation}
Here $x^\pm = \frac{x^0 \pm x^3}{\sqrt{2}}$, $z$ is the coordinate
describing the 5th dimension such that the boundary of the AdS space
is at $z=0$, and $L$ is the radius of $S_5$.  According to holographic
renormalization \cite{deHaro:2000xn}, $\langle T_{--} (x^-) \rangle$
is the expectation value of the energy-momentum tensor for a single
ultrarelativistic nucleus moving along the light-cone in the
$x^+$-direction in the gauge theory. We assume that the nucleus is
made out of nucleons consisting of $N_c^2$ ``valence gluons'' each,
such that $\langle T_{--} (x^-) \rangle \propto N_c^2$, and the metric
\peq{1nuc} has no $N_c^2$-suppressed terms in it.

\begin{figure}[h]
  \begin{center}
    \includegraphics[width=3cm]{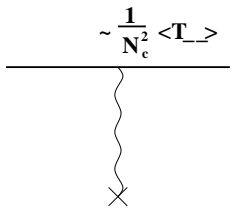}
  \end{center}
  \caption{A representation of the metric (\protect\ref{1nuc}) as a graviton 
    (wavy line) exchange between the nucleus at the boundary of AdS
    space (the solid line) and the point in the bulk where the metric
    is measured (denoted by a cross). }
  \label{1gr}
\end{figure}

The metric in \eq{1nuc} is an exact solution of Einstein equations in
AdS$_5$: 
\begin{align}
  \label{ein}
  R_{\mu\nu} + \frac{4}{L^2} \, g_{\mu\nu} = 0.
\end{align}
It can also be represented perturbatively as a single graviton
exchange between the source nucleus at the AdS boundary and the
location in the bulk where we measure the metric/graviton field. This
is shown in \fig{1gr}, where the solid line represents the nucleus and
the wavy line is the graviton propagator. Incidentally a single
graviton exchange, while being a first-order perturbation of the empty
AdS space, is also an exact solution of Einstein equations. This means
higher order tree-level graviton diagrams are zero (cf. classical
gluon field of a single nucleus in covariant gauge in the Color Glass
Condensate (CGC) formalism \cite{Kovchegov:1997pc}).

\begin{figure}
  \begin{center}
    \includegraphics[width=12cm]{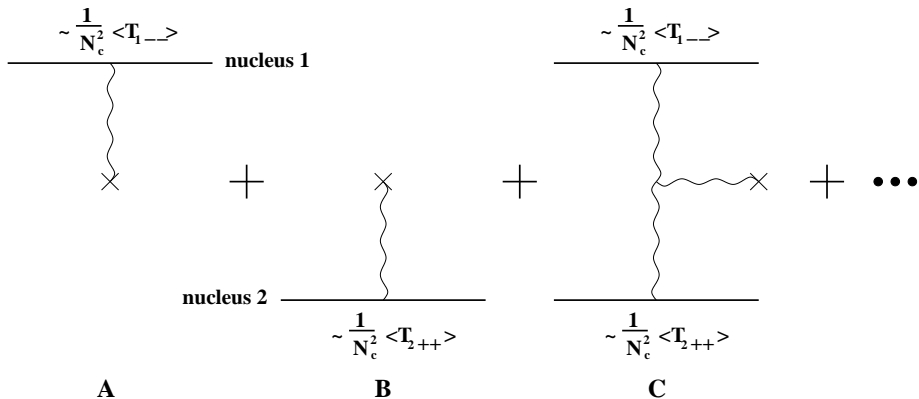}
  \end{center}
  \caption{Diagrammatic representation of the metric in \protect\eq{2nuc1}. 
    Wavy lines are graviton propagators between the boundary of the
    AdS space and the bulk.  Graphs A and B correspond to the metrics
    of the first and the second nucleus correspondingly.  Diagram C is
    an example of the higher order graviton exchange corrections.}
  \label{pert}
\end{figure}

Now let us try to find the geometry dual to a collision of two shock
waves with the metrics like that in \eq{1nuc}.  Defining $ t_1 (x^-)
\, \equiv \, \frac{2 \, \pi^2}{N_c^2} \, \langle T_{1 \, --} (x^-)
\rangle$ and $t_2 (x^+) \, \equiv \, \frac{2 \, \pi^2}{N_c^2} \,
\langle T_{2 \, ++} (x^+) \rangle$ for the energy-momentum tensors of
the two shock waves in the boundary theory, we write the metric
resulting from such a collision as
\begin{eqnarray}\label{2nuc1}
  ds^2 \, = \, \frac{L^2}{z^2} \, \bigg\{ -2 \, dx^+ \, dx^- + d
  x_\perp^2 + d z^2 + t_1 (x^-) \, z^4 \, d x^{- \, 2}  
  + t_2 (x^+) \, z^4 \, d
  x^{+ \, 2} \nonumber \\ + \, \mbox{higher order graviton exchanges}
  \bigg\}
\end{eqnarray}
The metric of \eq{2nuc1} is illustrated in \fig{pert}. The first two
terms in \fig{pert} (diagrams A and B) correspond to one-graviton
exchanges which constitute the individual metrics of each of the
nuclei, as shown in \eq{1nuc}. Our goal below is to calculate
higher-order corrections to these terms, which are illustrated by the
diagram C in \fig{pert} and by the ellipsis following it. \fig{pert}
illustrates that construction of dual geometry to a shock wave
collision in AdS$_5$ consists of summing up all tree-level graviton
exchange diagrams, similar diagrammatically to the classical gluon
field formed by heavy ion collisions in CGC
\cite{Kovner:1995ts,Kovchegov:2000hz}.  While classical gluon fields
lead to free-streaming final state \cite{Krasnitz:2002mn}, as we will
argue below, their AdS graviton ``dual'' will lead to an ideal
hydrodynamic final state for the gauge theory similar to the one found
in \cite{Janik:2005zt}, though at the same time different from
\cite{Janik:2005zt} due to non-trivial rapidity dependence in the case
at hand.


\section{Perturbative Solution of Einstein Equations}
\label{gensol}

We begin by parametrizing the metric as
\begin{align}\label{pA_gen}
  ds^2 \, = \, \frac{L^2}{z^2} \, \bigg\{ -\left[ 2 + g (x^+, x^-, z)
  \right] \, dx^+ \, dx^- + \left[ t_1 (x^-) \, z^4 + f (x^+, x^-, z)
  \right] \, d x^{- \, 2} \notag \\ + \left[ t_2 (x^+) \, z^4 +
    {\tilde f} (x^+, x^-, z) \right] \, d x^{+ \, 2} + \left[ 1 + h
    (x^+, x^-, z) \right] \, d x_\perp^2 + d z^2 \bigg\},
\end{align}
where the functions $f$, $\tilde f$, $g$, and $h$ are zero before the
collision, i.e., for either $x^+ < 0$ and/or $x^- <0$. The exact
Einstein equations \peq{ein} for $f$, $\tilde f$, $g$, and $h$ are
rather complicated and are not going to be presented here. Instead we
are going to solve Einstein equations perturbatively.

To be more specific let us consider in the boundary theory a collision
of two ultrarelativistic nuclei with large light-cone momenta per
nucleon $p_1^+$, $p_2^-$, and atomic numbers $A_1$ and $A_2$. Writing
\begin{align}\label{deltas}
  t_1 (x^-) = \mu_1 \, \delta (x^-), \ \ \ t_2 (x^+) = \mu_2 \, \delta
  (x^+)
\end{align}
we want to expand Einstein equations in the powers of $\mu_1$ and
$\mu_2$ \cite{Grumiller:2008va,Albacete:2008vs}. The two scales
$\mu_1$ and $\mu_2$ can be expressed terms of physical parameters in
the problem \cite{Albacete:2008vs,Albacete:2008ze}
\begin{align}\label{mus}
  \mu_{1} \sim p_{1}^+ \, \Lambda_1^2 \, A_1^{1/3}, \ \ \ \mu_{2} \sim
  p_{2}^- \, \Lambda_2^2 \, A_2^{1/3}.
\end{align}
$\Lambda_1$ and $\Lambda_2$ are the typical transverse momentum scales
describing the two nuclei \cite{Albacete:2008vs}, similar to the
saturation scales. Note that $\mu_1$ and $\mu_2$ are independent of
$N_c$. As follows from a simple dimensional analysis, combined with
Lorentz-transformation properties of the relevant quantities, the
expansion parameters in four dimensions would be
\begin{align}
  \label{params}
  \mu_1 \, ( x^- )^2 \, x^+, \ \ \ \mbox{and} \ \ \ \mu_2 \, ( x^+ )^2
  \, x^-,
\end{align}
such that the expansion is valid only at early times, when these
parameters are small.

Linearizing Einstein equations \peq{ein} in $f$, $\tilde f$, $g$, and
$h$ we solve the obtained system of differential equation to obtain
\cite{Albacete:2008vs}
\begin{equation}\label{hsol}
  h (x^+, x^-, z) \, = \, h_0 (x^+, x^-) \, z^4 + h_1 (x^+, x^-) \,
  z^6
\end{equation}
where $h_0$ and $h_1$ are determined by the causal solutions of the
following equations
\begin{equation}\label{h0eq}
  (\partial_+ \, \partial_-)^2 \, h_{0} (x^+, x^-) \, = \, 8 \, t_1
  (x^-) \, t_2 (x^+),
\end{equation}
\begin{equation}\label{h1eq2}
  \partial_+ \, \partial_- \, h_1 (x^+, x^-) \, = \, \frac{4}{3} \,
  t_1 (x^-) \, t_2 (x^+).
\end{equation}
$f$, $\tilde f$, and $g$ are easily expressed in terms of $h (x^+,
x^-, z)$ from \eq{hsol} (see \cite{Albacete:2008vs}).

This lowest-order perturbative solution leads to the energy density of
the produced medium (in the center-of-mass frame)
\cite{Grumiller:2008va,Albacete:2008vs}
\begin{align}
  \label{energy}
  \epsilon (\tau) \, = \, \frac{N_c^2}{\pi^2} \, \mu_1 \, \mu_2 \,
  \tau^2
\end{align}
where $\tau = \sqrt{2 \, x^+ \, x^-}$. The corresponding
center-of-mass energy-momentum tensor is
\begin{equation}\label{emt_ads}
  \langle T^{\mu\nu} \rangle \, = \,
  \left( \begin{array}{cccc} \epsilon (\tau) & 0 & 0 & 0 \\
      0 & 2 \, \epsilon  (\tau) & 0 & 0 \\
      0 & 0 & 2 \, \epsilon (\tau) & 0  \\
      0 & 0 & 0 & - 3 \, \epsilon (\tau) \end{array} \right)
\end{equation}
in terms of the $x^0, x^1, x^2, x^3$ components. One can see that the
longitudinal pressure component in \eq{emt_ads} is large and negative.
Under boosts the $T^{00}$ and $T^{33}$ components of the
energy-momentum tensor mix. This implies that there is a frame in which
the energy density is negative $T'_{00} < 0$.  At this point it is not
clear whether this result presents a problem, as there may be nothing
wrong with energy density becoming negative for a short period of time
\cite{Grumiller:2008va}. As we will see below, further evolution of
the energy density with time leads to disappearance of this negative
energy-density problem.

Higher-order perturbative corrections to the energy-momentum tensor
\peq{emt_ads} in powers of $\mu_1$ and $\mu_2$ have been found in
\cite{Grumiller:2008va,Albacete:2008vs} up to the fourth order in
$\mu_i$ (i.e., up to $O(\mu_1^3 \mu_2), O(\mu_1^2 \mu_2^2),O(\mu_1
\mu_2^3)$).


\section{All-Order Resummation in $\mu_2$}

\begin{figure}[h]
\begin{center}
\epsfxsize=6cm
\leavevmode
\hbox{\epsffile{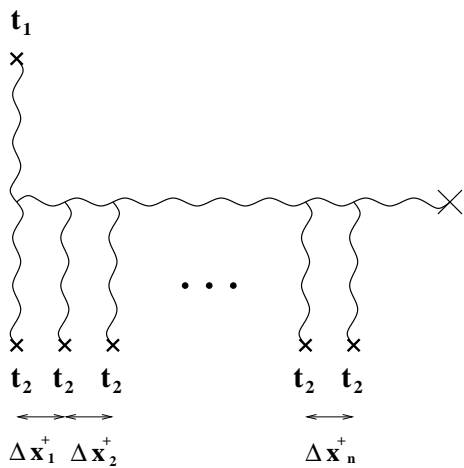}}
\end{center}
\caption{A diagram contributing to the metric of an asymmetric 
collision of two shock waves.}
\label{pAfig}
\end{figure}

The exact solution of Einstein equations for the collision of two
shock waves involves resummation of both parameters in \eq{params} to
all orders: such calculation appears to be very hard to do. Instead
one can resum all orders of $\mu_2 \, ( x^+ )^2 \, x^-$ while keeping
at the lowest order in $\mu_1 \, ( x^- )^2 \, x^+$. The corresponding
diagram is shown in \fig{pAfig}: in it one resums multiple graviton
exchanges with one shock wave ($t_2$), while exchanging only one
graviton with the other shock wave ($t_1$). By analogy with
perturbative QCD calculations we will refer to these class of diagrams
as to the ``proton-nucleus'' scattering, with the shock wave $t_1$
being the proton and the shock wave $t_2$ the nucleus. The
applicability region of such approximation is defined by
\begin{align}\label{bounds}
  \mu_1 \, (x^-)^2 \, x^+ \, \ll \, 1, \ \ \ \mu_2 \, (x^+)^2 \, x^-
  \sim 1,
\end{align}
which shows that the resummation is applicable to nucleus-nucleus
collisions, only in the part of the forward light-cone defined by
\eq{bounds}.

Such resummation was performed in \cite{Albacete:2009ji} using the
eikonal approximation. The result for the expectation value of the
energy-momentum tensor reads \cite{Albacete:2009ji}
\begin{subequations}\label{Tmnd}
\begin{align}
  \langle T^{++}\rangle \, & = \, - \frac{N_c^2}{2 \, \pi^2} \,
  \frac{4 \, \mu_1 \, \mu_2 \, (x^+)^2 \, \theta (x^+) \, \theta
    (x^-)}{\left[ 1 + 8 \, \mu_2 \, (x^+)^2 \, x^- \right]^{3/2}}, \label{T++} \\
  \langle T^{--}\rangle \, & = \, \frac{N_c^2}{2 \, \pi^2} \, \theta
  (x^+) \, \theta (x^-) \, \frac{\mu_1}{2 \, \mu_2 \, (x^+)^4} \,
  \frac{1}{\left[ 1 + 8 \, \mu_2 \, (x^+)^2 \, x^- \right]^{3/2}}
  \notag \\ & \times \, \bigg[ 3 - 3 \, \sqrt{1 + 8 \, \mu_2 \,
    (x^+)^2 \, x^-} + 4 \, \mu_2 \, (x^+)^2 \, x^- \notag \\ & \times
  \, \left( 9 + 16 \, \mu_2 \, (x^+)^2 \, x^- - 6 \, \sqrt{1 + 8 \,
      \mu_2 \, (x^+)^2 \, x^-}
  \right) \bigg] , \\
  \langle T^{+-}\rangle \, & = \, \frac{N_c^2}{2 \, \pi^2} \, \frac{8
    \, \mu_1 \, \mu_2 \, x^+ \, x^- \, \theta (x^+) \, \theta
    (x^-)}{\left[ 1 + 8 \, \mu_2 \, (x^+)^2 \, x^- \right]^{3/2}}, \\
  \langle T^{\, i j}\rangle \, & = \, \delta^{ij} \, \frac{N_c^2}{2
    \,\pi^2} \, \frac{8 \, \mu_1 \, \mu_2 \, x^+ \, x^- \, \theta
    (x^+) \, \theta (x^-)}{\left[ 1 + 8 \, \mu_2 \, (x^+)^2 \, x^-
    \right]^{3/2}}.
\end{align}
\end{subequations}
Provided the complexity of the problem at hand, the resulting formulas
(\ref{Tmnd}) for the energy-momentum tensor are remarkably simple!

Now we can ask a question: what kind of medium is produced in these
strongly coupled proton-nucleus collisions? Is it described by ideal
hydrodynamics, just like Bjorken hydrodynamics was obtained in
\cite{Janik:2005zt}? In our case the produced matter distribution is
obviously rapidity-dependent, so it is slightly more tricky to check
whether Eqs. (\ref{Tmnd}) constitute an ideal hydrodynamics, i.e.,
whether it can be written as
\begin{align}\label{hydro}
  T^{\mu\nu} \, = \, (\epsilon + p) \, u^\mu \, u^\nu - p \,
  \eta^{\mu\nu}
\end{align}
with the positive energy density $\epsilon$ and pressure $p$.
$\eta^{\mu\nu}$ is the metric of the four-dimensional Minkowski
space-time and $u^\mu$ is the fluid 4-velocity.

For the particular case at hand it is easy to see that the
energy-momentum tensor in \eq{Tmnd} can not be cast in the ideal
hydrodynamics form of (\ref{hydro}). In the case of ideal
hydrodynamics one has
\begin{align}
  T^{++} \, = \, (\epsilon + p) \, (u^+)^2 \, > \, 0.
\end{align}
At the same time $\langle T^{++}\rangle$ in \eq{T++} is {\sl negative
  definite}. Therefore the ideal hydrodynamic description is not
achieved in the proton-nucleus collisions. We believe this result is
due to limitations of this proton-nucleus approximation. As we will
show below, thermalization (black hole production) is inevitable in
the collision of the two shock waves considered here. Our conclusion
is then that thermalization/isotropization of the medium does not
happen in the space-time region defined by the bounds in \eq{bounds}.
What we found in \eq{Tmnd} is a medium at some intermediate stage, on
its way to thermalization at a later time.  It appears that one needs
to solve the full nucleus-nucleus scattering problem to all orders in
both $\mu_1$ and $\mu_2$, to obtain a medium described by ideal
hydrodynamics.


\section{Stopping of Nuclei After Collision}

\begin{figure}
  \begin{center}
    \includegraphics[width=7cm]{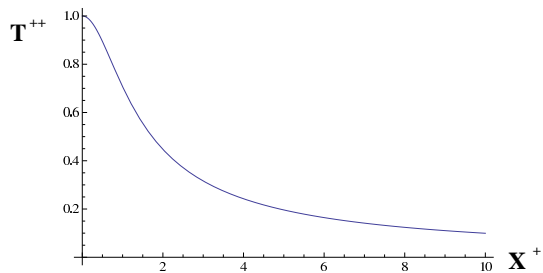}
  \end{center}
  \caption{$T^{++}$ component of the proton's energy-momentum tensor 
    after the collision as a function of light cone time $x^+$
    (arbitrary units).}
  \label{pstop}
\end{figure}

To better understand dynamics of the shock wave collisions let us
follow one of the shock waves after the interaction. First we
``smear'' the delta-function profile of that shock wave:
\begin{equation}\label{t1s}
  t_1 (x^-) \, = \, \frac{\mu_1}{a_1} \, \theta (x^-) \,
  \theta (a_1 - x^-).
\end{equation}
Here $a_1 \, \propto \, R_1 \, \frac{\Lambda_1}{p_1^+} \, \propto \,
\frac{A_1^{1/3}}{p_1^+}$, where the nucleus of radius $R_1$ has $A_1$
nucleons in it.  The ``$+ +$'' component of the energy-momentum tensor
of a shock wave after the collision at $x^- = a_1 /2$ is
\cite{Albacete:2008vs}
\begin{equation}\label{stop}
  \langle T^{+ \, +} (x^+ , x^- = a_1 /2) \rangle \, = \, 
\frac{N_c^2}{2 \, \pi^2} \,
  \frac{\mu_1}{a_1} \left[ 1 - 2 \, \mu_2 \, x^{+\, 2} \, a_1 \right].
\end{equation}
The first term on the right of \eq{stop} is due to the original shock
wave while the second term describes energy loss due to graviton
emission. \eq{stop} shows that $\langle T^{+ \, +} \rangle$ of a
nucleus becomes {\sl zero} at light-cone times (as $p_1^+ \approx
p_2^-$ in the center-of-mass frame)
\begin{equation}\label{stoptime}
  x^+_{stop} \, \sim \, \frac{1}{\sqrt{\mu_2 \, a_1}} \, \sim \, 
\frac{1}{\Lambda_2 \, A_1^{1/6} \, A_2^{1/6}}. 
\end{equation}
Zero $\langle T^{+ \, +} \rangle$ would mean {\sl stopping} of the
shock wave and the corresponding nucleus. The result can be better
understood by doing all-order resummation of graviton exchanges with
one shock wave performed above for ``proton-nucleus collisions''
\cite{Albacete:2009ji}. The full result for the proton's ``$+ +$''
component of the energy-momentum tensor is
\begin{equation}\label{stop_pA}
  \langle T^{++} \rangle \, = \, \frac{N_c^2}{2 \, \pi^2} \,
  \frac{\mu_1}{a_1} \, \frac{1}{\sqrt{1 + 8 \, \mu_2 \, (x^+)^2 \,
      x^-}}, \ \ \ \mbox{for} \ \ \ 0 < x^- < a_1.
\end{equation}
\eq{stop_pA} is illustrated in \fig{pstop}, in which one can see that
the proton loses all of its light cone momentum over a rather short
time.

We thus conclude that the collision of two nuclei at strong coupling
leads to a necessary stopping of the two nuclei shortly after the
collision. If the nuclei stop completely in the collision, the strong
interactions between them are almost certain to thermalize the system,
probably leading to Landau hydrodynamics \cite{Landau:1953gs}.
(Rapidity-independent Bjorken hydrodynamics \cite{Bjorken:1982qr}
seems to be unlikely after stopping. Even before stopping the
energy-momentum tensor in \eq{Tmnd} is strongly rapidity-dependent.)


\section{Thermalization: Trapped Surface Analysis}

While the exact solution of Einstein equations for the colliding shock
waves remains elusive, one can infer whether a black hole will be
created in the bulk following such collisions by performing a trapped
surface analysis \cite{Penrose,Eardley:2002re}. A trapped surface
analysis for shock waves with sources in the bulk has been carried out
before in \cite{Gubser:2008pc,Lin:2009pn,Gubser:2009sx}. However, with
the resulting trapped surface being centered around the sources, it is
not clear to what extend the trapped surface is the property of these
bulk sources, and what happens to the trapped surface when there are
no bulk sources, as is the case for our shock wave \peq{1nuc}.

Performing a trapped surface analysis for a shock wave without sources
\peq{1nuc} does not allow one to uniquely fix the position of the
trapped surface \cite{Kovchegov:2009du}. We therefore start with a
shock wave having an extended source in the bulk with the only
non-zero component of the bulk energy-momentum tensor being $J_{--} \,
= \, \frac{E}{z_0 \, L} \, \delta (x^-) \, \delta (z - z_0)$. The
corresponding metric is \cite{Lin:2009pn}
\begin{align}\label{nuc1s}
  ds^2 \, = \, \frac{L^2}{z^2} \, \left\{ -2 \, dx^+ \, dx^- + \phi
    (z) \, \delta (x^-) \, d x^{- \, 2} + d x_\perp^2 + d z^2
  \right\}
\end{align}
with 
\begin{align}\label{phi_sol}
  \phi (z) \, = \, \frac{2 \, \pi^2 \, E \, L^2}{N_c^2} \, \left\{
    \begin{array}{c}
      \frac{z^4}{z_0^4}, \ z \le z_0 \\~\\
        1, \ z > z_0.
    \end{array}
\right.
\end{align}
As one can readily see, the metric \peq{nuc1s} reduces to that in
\eq{1nuc} if we send the source to the IR infinity in the bulk by
taking $z_0 \rightarrow \infty$ limit of the metric in Eqs.
(\ref{nuc1s}) and (\ref{phi_sol}) keeping $\mu$ defined by
\begin{align}\label{mu}
  \mu \, = \, \frac{2 \, \pi^2}{N_c^2} \, \frac{E \, L^2}{z_0^4}
\end{align}
fixed \cite{Kovchegov:2009du}.

\begin{figure}[h]
\begin{center}
\epsfxsize=9cm
\leavevmode
\hbox{\epsffile{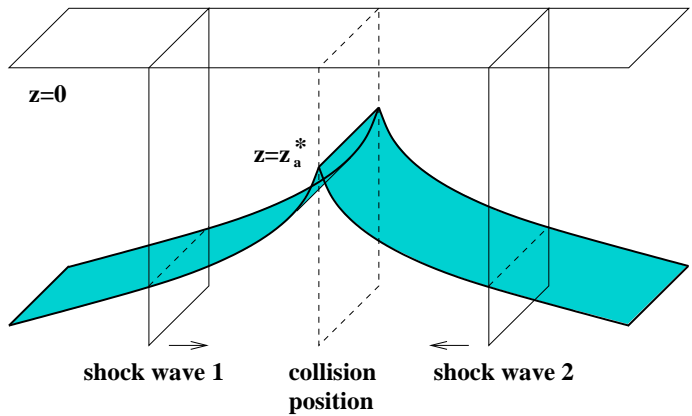}}
\end{center}
\caption{An illustration of the trapped surface in the collision of 
  two sourceless shock waves. Vertical axis is the bulk $z$-direction,
  the horizontal left-right axis can be thought of either as the
  collision axis or as the time direction.  The trapped surface is
  shaded.}
\label{surf}
\end{figure}

Marginal trapped surface for a collision of two shock waves given by
\eq{nuc1s} was found in \cite{Lin:2009pn}. In the $z_0 \rightarrow
\infty$, $\mu$ fixed limit that trapped surface reduces to
\begin{align}\label{fintr}
  x^+ \, = \,0, \ x^- \, = \, - \frac{\mu}{2} \, \left[ z^4 -
    \mu^{-4/3} \, 2^{-2/3} \right]
\end{align}
with an analogous expression for the other shock wave obtained by
interchanging $x^+ \leftrightarrow x^-$ in \eq{fintr}. (For simplicity
we work in the center-of-mass frame where $\mu_1 = \mu_2 = \mu$.) The
trapped surface in \eq{fintr} is independent of the shape of the
source being sent to deep IR, as shown in \cite{Kovchegov:2009du}. The
trapped surface from \eq{fintr} is depicted in \fig{surf}.

The existence of trapped surface proves that gravitational collapse is
inevitable. That is a black hole will be created in a bulk following a
collision of two sourceless shock waves. In the boundary theory this
means that a thermal medium is created, which is described by ideal
hydrodynamics.

The (lower bound for the) produced entropy per unit transverse area
$A_\perp$ can be found by calculating the area of the trapped surface,
which yields \cite{Kovchegov:2009du}
\begin{align}
  \frac{S}{A_\perp} \, = \, \frac{N_c^2}{2 \, \pi^2} \, \left( 2 \,
    \mu_1 \, \mu_2 \right)^{1/3}.
\end{align}
Since the trapped surface analysis does not ``know'' anything about
shock wave thickness (e.g. $a_1$), we conclude that the thermalization
time is only a function of $\mu_1$ and $\mu_2$, which gives
\begin{align}
  \label{thermtime}
  \tau_{th} \, \sim \, \frac{1}{(\mu_1 \, \mu_2)^{1/6}},
\end{align}
in agreement with the thermalization time suggested in
\cite{Grumiller:2008va}. While numerically this thermalization time is
too short to be relevant for RHIC data, it is parametrically shorter
than the stopping time \peq{stoptime}, making our model somewhat more
relevant for description of real-life collisions. As $\mu_1 \, \mu_2
\sim s$ with $s$ the center of mass energy of the collision, we get
\begin{align}
  \frac{S}{A_\perp} \, \propto \, s^{1/3}
\end{align}
in agreement with the result obtained in \cite{Gubser:2008pc}. Finally
note that, since, as follows from \eq{mus}, $\mu_1$ and $\mu_2$ are
$N_c$-independent, the produced entropy scales $\propto N_c^2$, in
agreement with $N_c$-counting in a perturbative QCD calculation of
particle production for a collision of two nuclei with $N_c^2$
``valence gluons'' in their nucleons.


\section*{Acknowledgements}

I am grateful to Javier Albacete, Shu Lin, and Anastasios Taliotis for
collaborating with me on various parts of this work
\cite{Albacete:2008vs,Albacete:2009ji,Kovchegov:2009du}.

This research is sponsored in part by the U.S. Department of Energy
under Grant No. {DE-SC0004286}.

%


\providecommand{\href}[2]{#2}\begingroup\raggedright

\endgroup

  

\end{document}